\documentstyle[aps,epsf,psfig,floats]{revtex}  
\begin{document}  
\draft
\widetext
\title{Physical origin of the buckling in CuO$_2$: Electron-phonon  
coupling and Raman spectra}  
\author{M. Opel and R. Hackl}  
\address{Walther Meissner Institut f\"ur Tieftemperaturforschung,  
Garching, D-85748, Germany}  
\author{T. P. Devereaux}  
\address{Department of Physics, The George Washington University,  
Washington, DC 20052}  
\author{A. Virosztek and A. Zawadowski}  
\address{Institute of Physics and Research Group of the Hungarian  
Academy  
of Sciences, Technical University of Budapest, H-1521  
Budapest, Hungary\\  
and Research Institute for Solid State Physics, P.O.Box 49, H-1525  
Budapest, Hungary}  
\author{A. Erb and E. Walker}  
\address{DPMC, Universit\'e de Gen\`eve, CH-1211 Gen\`eve,  
Switzerland}  
\author{H. Berger and L. Forr\'o}  
\address{EPFL, Ecublens, CH-1015 Lausanne, Switzerland}  
\date{\today}  
\maketitle  
  
\widetext  
\begin{abstract}  
It is shown theoretically that the buckling of the CuO$_{2}$   
planes in certain cuprate systems can be explained in   
terms of an electric field across the planes which   
originates from different valences of atoms above and   
below the plane. This field results also in a   
strong coupling of the Raman-active out-of-phase vibration of the   
oxygen atoms ($B_{1g}$ mode) to the electronic charge transfer 
between the two oxygens in the CuO$_{2}$ plane. Consequently, the electric 
field can be deduced from the Fano-type line shape of the $B_{1g}$ 
phonon. Using the electric field estimated from the electron-phonon coupling 
the amplitude of the buckling is calculated and found to be in good agreement 
with the structural data. Direct experimental support for the idea   
proposed is obtained in studies of YBa$_{2}$Cu$_{3}$O$_{6+x}$ and   
Bi$_{2}$Sr$_{2}$(Ca$_{1-x}$Y$_{x}$)Cu$_{2}$O$_{8}$ with  
different oxygen and   
yttrium doping, respectively, including antiferromagnetic samples.   
In the latter compound, symmetry breaking by replacing Ca partially 
by Y leads   
to an enhancement of the electron-phonon coupling by   
an order of magnitude. 
\end{abstract}  
\widetext  
  
\section{introduction}  
The physical origin of the buckling of the CuO$_2$ plane in certain high  
temperature superconductors, where the oxygen atoms are placed outside  
of the plane of the copper atoms, is a long standing problem. There are two  
possible explanations: (i) the surroundings of the CuO$_2$ plane is not  
symmetrical, thus an electric field perpendicular to the  
plane acting on atoms with different charges deforms that plane, and  
(ii) the energy associated with chemical bonds (especially the bond  
between the oxygen $\pi$-orbitals perpendicular to the plane and the  
coppers) results in a spontaneous symmetry breaking.   
The present paper provides direct experimental evidence to  
support the first possibility in accordance with the theory.  
  
It has been suggested previously\cite{DVZ1} that in the presence of  
a perpendicular electric field a charge transfer between the two different 
oxygens O(2) and O(3)   
invokes a deformation of the type of the $B_{1g}$ Raman-active phonon  
at 330cm$^{-1}$, where the two oxygens move in an out of phase  
manner perpendicular to the plane (Fig. 1a). 
The resulting electron-phonon  
coupling was then used to interpret the Fano interference in the Raman 
spectra of YBa$_{2}$Cu$_{3}$O$_{6+x}$ (Y-123), where  
the light is scattered by both the $B_{1g}$ phonon and the electrons involved  
in the charge transfer in the plane described above. For a  
CuO$_2$ plane in a more symmetrical environment (where the electric  
field is negligible) such a linear coupling does not exist since 
the energy correction due to the deformation of the chemical bonds 
is at least quadratic in the displacement of the oxygens. Hence,  
only much weaker two-phonon processes contribute.   
  
The coupling strength and, consequently, the value of the electric field  
perpendicular to the plane can be determined from the Fano line shape of the  
$B_{1g}$ phonon using the expressions derived in Ref. \cite{DVZ1}. In  
order to follow closely the doping dependence of the line shape we will use  
here, in contrast to \cite{DVZ1}, a more realistic
three band model for the electronic  
system. It will be described in some detail thus augmenting an earlier  
publication which gave just a brief summary \cite{DVZ3}. The Raman- 
active continuum which interferes with the $B_{1g}$ phonon is assumed  
to originate from strongly relaxing electrons in a single conduction  
band such as proposed in Refs. \cite{nfl,ZC,NAFL} and will be introduced  
phenomenologically as in \cite{DVZ1}. Recently, Sherman and  
coworkers considered the coupling of the $B_{1g}$ phonon to non-
interacting Bloch electrons including intra and inter-band intermediate  
states in order to understand the phonon intensity  
\cite{sherman}. Here, the focus is placed on the interference of the  
phonon and the electronic continuum which are excited by the light through
two initially independent channels. In our approach the direct coupling of 
the phonon to the light is parameterized and not explicitly considered. 
Our goal is to better understand the interaction between the 
phonon and the conduction electrons. Therefore, both Y-123, where the  
surroundings of the CuO$_2$ plane is highly asymmetric, and  
Bi$_{2}$Sr$_{2}$CaCu$_{2}$O$_{8}$ (Bi-2212),  
where the plane is placed in a more symmetrical position with respect  
to the charges of the nearby atoms, will be studied. 

\begin{figure}  
\vskip 0cm  
\epsfxsize=16cm  
\epsffile{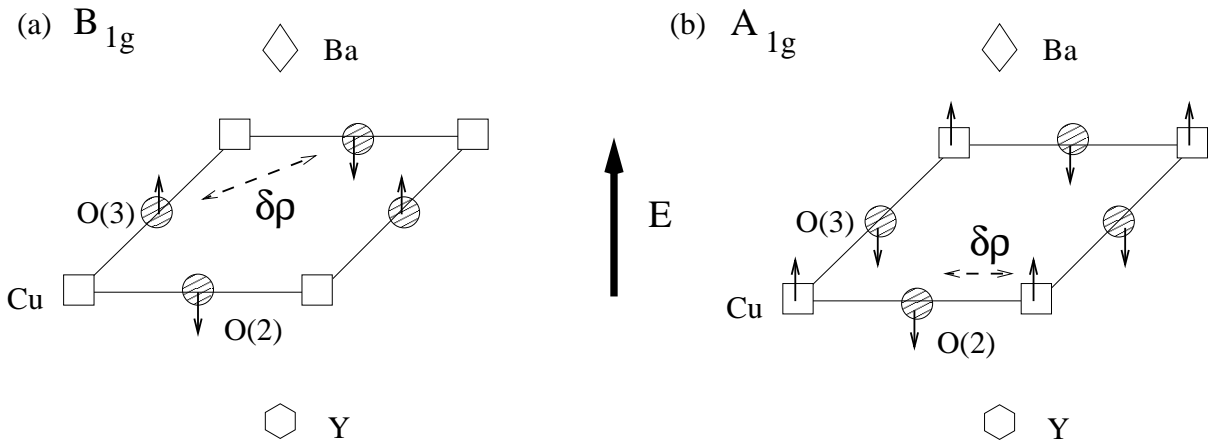}  
\vskip 0.1cm  
\caption{The unit cell of the CuO$_{2}$ plane in Y-123 shown 
with the atomic   
displacements corresponding to the $B_{1g}$ (a) and $A_{1g}$ (b)  
phonons. The electric field $E$ perpendicular to the planes is due  
to the asymmetric environment (Ba$^{2+}$ above, Y$^{3+}$ below),  
and causes a static  
deformation of the $A_{1g}$ type, usually referred to as buckling  
(also called dimpling as in Ref.\ \protect\onlinecite{Andersen}).   
$\delta\rho$ and the double arrows denote the charge transfer which  
accompanies the lattice vibration. In Bi-2212, Ba$^{2+}$ is replaced by  
Sr$^{2+}$ and, at optimal doping, Y$^{3+}$ by Ca$^{2+}$. 
Then the electrical field $E$ is negligible. }  
\label{phonons}  
\end{figure}  
  
The $B_{1g}$ phonon in metallic cuprate systems has actually been
studied intensively by several groups 
\cite{DVZ3,KL,Y123old,alt}. The main emphasis was placed on the 
anomalies at the transition to the superconducting state. It was 
shown, for instance, that the change in frequency of the $B_{1g}$ 
phonon at $T_{c}$ vanishes already at moderate 
underdoping\cite{alt} pointing at a considerable change 
in the electron-phonon coupling. While the phonon 
renormalization at $T_{c}$ has been the subject of numerous theoretical
treatments\cite{Tctheory}, 
no attempt has been made yet to link the doping dependence 
of the coupling to a realistic band structure, i.e. a microscopic model. 
For a comparison with the model predictions, a systematic experimental 
study of the phonon covering the complete doping range from 
antiferromagnetic (AF) insulating to slightly overdoping in comparable, 
high-quality samples 
is essential. Therefore, we present  
not only data on differently doped metallic samples but also new data on  
undoped antiferromagnetic ones in order to get an idea of the intrinsic line  
width and to study in detail the influence of a reduction of the continuum  
scattering on the line shape of the phonon. 
 
As already shown earlier \cite{DVZ3,KL} the $B_{1g}$ phonon has a  
reasonably Lorentzian line shape for the symmetric compound Bi-2212  
indicating the lack of any substantial coupling between that phonon and  
the charge transfer between  
the two oxygens in accordance with the absence of an electric field.  
This striking difference to the data of Y-123 \cite{Y123old}   
can be considered as strong evidence in favor of the linear (involving  
one phonon) electron-phonon coupling due to the electric field.  
In order to complete this argument, we gradually replace Ca$^{2+}$ by  
Y$^{3+}$ in the Bi-2212 compound  
where the doping most likely breaks locally the approximate  
reflection symmetry  
through the CuO$_2$ plane, thereby making this material similar to 
Y-123. In fact, the  phonon becomes more asymmetric and the 
electron-phonon coupling increases by an order of 
magnitude\cite{DVZ3}. However, this increase does not at all result 
in an enhancement of the superconducting transition temperature: 
$T_c$ is rather reduced on substituting Ca with Y and follows the usual 
$T_c$ vs. doping curve.
  
Finally, the value for the electron-phonon coupling obtained from the  
analysis of the line shape is found to be in excellent agreement with  
the value calculated from the magnitude of the electric field due to the  
experimentally observed buckling\cite{Jorg} and the restoring force   
calculated from the frequency of the $A_{1g}$ phonon (Fig. 1b).  
The coupling depends on changes in the doping via the proximity of the
van Hove singularity in the density of states to the Fermi level $E_{F}$.  
  
The paper is organized as follows: In section II the theoretical model is  
described in some detail and the relevant expressions are derived. In  
section III and IV experimental details and results are being given.  
Finally, the consequences are discussed.   
  
\section{Theoretical Model}  
  
Much of the development of the crystal field model was given in  
Ref. \cite{DVZ1}, and details of the calculations are found therein.   
We start with this and only  
point out where changes are made.   
The model constructed in Ref. \cite{DVZ1} started from a three band  
model  
where only the Cu-O hopping was taken into consideration.   
However, one must  
add O-O hopping in order to represent the curvature and centering of  
the  
Fermi surface as revealed from angle-resolved photo-emission  
experiments.  
Therefore we augment the results from \cite{DVZ1} to include direct  
O-O hopping. We  
start by first considering a simple 3 band model for the CuO$_{2}$ plane  
with Cu-O hopping amplitude $t$ and O-O hopping amplitude  
$t^{\prime}$:  
\begin{equation}  
H_{0}=\varepsilon\sum_{{\bf n},\sigma}b^{\dagger}_{{\bf n},\sigma}  
b_{{\bf n},\sigma}  
+t\sum_{{\bf n},\sigma,\pmb{$\delta$}}[P_{\pmb{$\delta$}}  
b^{\dagger}_{{\bf n},\sigma}  
a_{{\bf n},\sigma,\pmb{$\delta$}}+h.c.]+  
t^{\prime}  
\sum_{{\bf  
n},\sigma}\sum_{\langle\pmb{$\delta$},\pmb{$\delta$}^\prime\rangle}  
P^\prime_{\pmb{$\delta$},\pmb{$\delta$}^\prime}a^\dagger_{{\bf n},  
\pmb{$\delta$},\sigma}a_{{\bf n},\pmb{$\delta$}^\prime,\sigma},  
\label{one}  
\end{equation}  
where $b^{\dagger}_{{\bf n},\sigma}$ creates an electron with spin  
$\sigma$ at a copper lattice site ${\bf n}$, while   
$a_{{\bf n},\sigma,\pmb{$\delta$}}$  
annihilates an electron at one of the neighboring oxygen sites  
${\bf n}+\pmb{$\delta$}/2$  
determined by the unit vector $\pmb{$\delta$}$ assuming the four values,  
$(\pm 1,0)$ and $(0,\pm 1)$. An oxygen atom between the two copper  
atoms at  
sites ${\bf n}$ and ${\bf n}+\pmb{$\delta$}$ is labeled by either  
$({\bf n},\pmb{$\delta$})$ or $({\bf n}+\pmb{$\delta$},- 
\pmb{$\delta$})$.  
As in \cite{DVZ1}, $\varepsilon=E_d-E_p$ is the difference of the Cu  
and O site  
energies and $P_{\pmb{$\delta$}}=\pm 1$  
depending on whether the orbitals (with real wavefunctions) have the  
same  
or opposite sign at the overlap region. Assuming Cu $d_{x^2-y^2}$ and  
O $p$ orbitals $P_{-\pmb{$\delta$}}=-P_{\pmb{$\delta$}}$, and we can  
choose  
$P_{(1,0)}=1$ and $P_{(0,1)}=-1$. Lastly,   
$P^{\prime}_{\pmb{$\delta$},\pmb{$\delta$}^\prime}$ denotes the  
overlap sign between an  
oxygen orbital at site ${\bf n}+\pmb{$\delta$}/2$ with a  
neighboring oxygen orbital at site ${\bf n}+\pmb{$\delta$}^\prime/2$.  
By our above  
convention these overlaps take the values $P^{\prime}_{{\bf x},{\bf y}}=  
P^{\prime}_{-{\bf x},-{\bf y}}=1,  
P^{\prime}_{{\bf x},-{\bf y}}=P^{\prime}_{-{\bf x},{\bf y}}=-1$,  
respectively. After Fourier transforming, the Hamiltonian now reads  
$H^{0}=\sum_{{\bf k},\sigma} H^0_{{\bf k},\sigma}$, where  
\begin{equation}  
H^0_{{\bf k},\sigma}=\varepsilon b^{\dagger}_{{\bf k},\sigma}b_{{\bf  
k},  
\sigma} +\{ ib^{\dagger}_{{\bf k},\sigma}[a_{x,{\bf k},\sigma}t_x({\bf  
k})  
-a_{y,{\bf k},\sigma}t_y({\bf k})]+h.c.\}+  
t^{\prime}_{\bf k}[a_{x,{\bf k},\sigma}^{\dagger}a_{y,{\bf  
k},\sigma}+h.c.],  
\label{two}  
\end{equation}  
with the prefactors  
\begin{equation}  
t_{\alpha}({\bf k})=2t\sin(ak_{\alpha}/2),~~~~~~~   
t^{\prime}({\bf k})=-4t^{\prime}\sin(ak_{x}/2)\sin(ak_{y}/2).  
\label{three}  
\end{equation}  
We can then easily diagonalize Eq. (\ref{two}) and obtain the 3 bands:  
\begin{equation}  
H^{0}_{{\bf k},\sigma}=\sum_{\beta} E_{\beta}({\bf k})  
d^{\dagger}_{\beta, {\bf k},  
\sigma}d_{\beta, {\bf k},\sigma},  
\label{four}  
\end{equation}  
where $\beta$ assumes the values $+,-,$ and $0$ for the anti-bonding,  
bonding, and non-bonding bands, respectively.   
The energies are then given by solving a third order root equation and  
are  
\begin{eqnarray}  
&&E_{+}({\bf k})=s_{+}({\bf k})+s_{-}({\bf k})+\varepsilon/3 \\  
&&E_{-}({\bf k})=-{1\over{2}}[s_{+}({\bf k})+s_{-}({\bf k})]+  
\varepsilon/3 +{i\sqrt{3}\over{2}}[s_{+}({\bf k})-s_{-}({\bf k})]   
\nonumber \\  
&&E_{0}({\bf k})=-{1\over{2}}[s_{+}({\bf k})+s_{-}({\bf k})]+  
\varepsilon/3 -{i\sqrt{3}\over{2}}[s_{+}({\bf k})-s_{-}({\bf k})]   
\label{five}  
\nonumber   
\end{eqnarray}  
with  
\begin{eqnarray}  
&&s_{\pm}({\bf k})=(r({\bf k})\pm\sqrt{q^{3}({\bf k})+r^{2}({\bf  
k})})^{1\over{3}}  
\\  
&&q({\bf k})=-{1\over{3}}[t_{x}^{2}({\bf k})+t_{y}^{2}({\bf k})  
+t^{\prime 2}({\bf k})]  
-\varepsilon^{2}/9 \nonumber \\  
&&r({\bf k})={\varepsilon\over{6}}[t_{x}^{2}({\bf k})+  
t_{y}^{2}({\bf k})-2t^{\prime 2}({\bf k})]-t^{\prime}({\bf k})  
t_{x}({\bf k})t_{y}({\bf k})  
+\varepsilon^{3}/27  
\label{six}  
\nonumber  
\end{eqnarray}  
Since $q^{3}({\bf k})+r^{2}({\bf k})$ is negative for all   
${\bf k}$, $s_{+}^{*}({\bf k})=s_{-}({\bf k})$   
and the energies are of course real.  
  
\begin{figure}  
\vskip  0cm  
\epsfxsize=8cm  
\epsffile{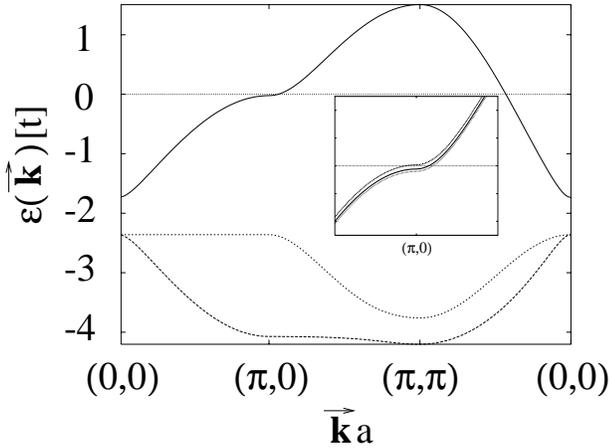}  
\vskip 0.1cm  
\caption{Band dispersion for the 3-band model with nearest and next -  
nearest - neighbor hopping for the parameters listed in the text  
corresponding to optimally doped YBa$_{2}$Cu$_{3}$O$_{7}$. The  
upper (middle,  
lower) curve corresponds to the anti-bonding (non-bonding, bonding)  
band,  
respectively. Inset shows the  
doping dependence for the upper band only in the vicinity of the  
van Hove point near $(\pi,0)$.  
The upper (middle,lower) curve represents over-(optimally, under-)doped  
systems, respectively.}  
\label{band}  
\end{figure}

The parameter values used,  
$\varepsilon=1$eV, $t=1.6$eV, and $t^{\prime}/t=0.35$, are  
similar to those used in Ref. \cite{Andersen}.  
The dispersion for the three bands is plotted for symmetry  
${\bf k}$ points along the Brillouin zone (BZ)  
in Fig. \ref{band} using these  
parameters and choosing the chemical potential such that the filling   
$\langle n \rangle=0.85$ for both spins. The upper (middle, lower)  
curve represents the band dispersion for the anti-bonding  
(non-bonding, bonding) band, respectively. Setting the  
direct O-O hopping integral $t^{\prime}=0$ would yield a  
dispersionless  
non-bonding band whose energy equaled the oxygen site energy, while the   
other bands recover the dispersion calculated in Ref. \cite{DVZ1}. The  
inset of Fig. \ref{band} shows the dispersion of the anti-bonding band in  
the  
vicinity of the $(\pi,0)$ point in the BZ evaluated for different fillings  
which we label as over-doped ($\langle n \rangle =0.8$), optimally-doped  
($\langle n \rangle =0.85$),  
and under-doped ($\langle n \rangle =0.875$), corresponding to the upper,  
middle, and lower  
curves in the inset, respectively. The filling changes the  
distance the flat part of the dispersion (van Hove point)  
is away from the Fermi level $E_{F}$,  
thus changing the density of states at $E_{F}$.  
Through this change in the density of states we attempt to  
understand the doping dependence of the coupling between the  
$B_{1g}$ optical phonon mode and the conduction electrons.  
  
At this point we can carry over the results from Ref. \cite{DVZ1}  
with little loss of generality. As in Ref. \cite{DVZ1}  
we consider only a reduced one band model  
appropriate for near half filling and take only the upper band   
$E_{+}({\bf k})$ into account.  
The electron - phonon reduced Hamiltonian  
for the $B_{1g}~~ {\bf q}=0$ phonon mode was given as  
\begin{equation}  
H_{el-ph}={1\over{\sqrt{N}}}\sum_{{\bf k},\sigma} g({\bf k})   
d^{\dagger}_{{\bf k},\sigma}d_{{\bf k},\sigma}[c+c^{\dagger}],  
\label{seven}  
\end{equation}  
where $N$ is the number of sites,
$d_{{\bf k},\sigma}$ annihilates an electron of spin $\sigma$ and  
momentum ${\bf k}$, and $c^{\dagger}$ creates a $B_{1g}$ phonon  
mode of  
wavevector ${\bf q}=0$. The   
coupling constant $g({\bf k})$   
of the $B_{1g}$ mode to an electron with momentum ${\bf k}$   
was evaluated in Ref. \cite{DVZ1} and is given by  
\begin{equation}  
g({\bf k})= eE\sqrt{{\hbar\over{2M\omega_{B_{1g}}}}}  
{1\over{\sqrt{2}}}  
[\mid\phi_{x}({\bf k})\mid^{2}-\mid\phi_{y}({\bf k})\mid^{2}],  
\label{eight}  
\end{equation}  
where $M$ is the oxygen mass, $\omega_{B_{1g}}$ is the phonon  
frequency,  
and $E$ is the ${\bf \hat z}$- component of the electric crystal field  
which for simplicity we take to be equal at both the O(2) and O(3) 
sites.  
The functions $\phi_{x,y}$ are the amplitudes of the oxygen wave  
functions in the tight-binding wave functions of the upper band which 
arise from diagonalizing  
the three band model. In the absence of O-O hopping, these functions  
were  
given in \cite{DVZ1}. Here, O-O hopping is included yielding
\begin{eqnarray}  
&&\phi_{x}({\bf k})={-i\over{N({\bf k})}} [E_{+}({\bf k}) t_{x}({\bf  
k})  
-t^{\prime}({\bf k})t_{y}({\bf k})], \nonumber \\  
&&\phi_{y}({\bf k})={i\over{N({\bf k})}} [E_{+}({\bf k}) t_{y}({\bf  
k})  
-t^{\prime}({\bf k})t_{x}({\bf k})], \nonumber \\  
&&{\rm with}~~~~~~  
N^{2}({\bf k})=[E_{+}^{2}({\bf k})-t^{\prime 2}({\bf k})]^{2}+  
[E_{+}({\bf k})t_{x}({\bf k})-t_{y}({\bf k})t^{\prime}({\bf k})]^{2}+  
[E_{+}({\bf k})t_{y}({\bf k})-t_{x}({\bf k})t^{\prime}({\bf k})]^{2}.  
\label{nine}  
\end{eqnarray}  
We now are in a position to calculate the full Raman response.  
Again we can utilize the results derived in Ref. \cite{DVZ1}.   
The generalized form of the Breit-Wigner or Fano lineshape  
describes the scattering of the light by the coupled electron-phonon  
system. The (independent) coupling strengths of the light to the phonon  
and to the electronic continuum with symmetry $\lambda$ will be denoted  
by $g_{p-p,\lambda}$ and $\gamma_{\lambda}$, respectively. The bare  
phonon when decoupled from the conduction electrons is characterized by  
the energy $\omega_{\lambda}$ and the intrinsic damping due to e.g.  
anharmonic phonon-phonon interaction by $\Gamma _{\lambda}^{i}$.  
The electronic continuum in symmetry channel $\lambda$ will be  
described by the electronic susceptibility $\chi_{\lambda}$. It is clear that  
it cannot originate from a non interacting gas formed by Bloch electrons;  
it rather should be traced back to strong renormalization effects such as  
electron-electron interaction \cite{nfl}, spin-fermion \cite{NAFL} 
or to scattering on static impurities  
\cite{ZC}. A parameterized form for $\chi_{\lambda}$ is used 
which has been derived by one of us when electron  
dynamics is considered in materials with nesting properties along the  
Fermi surface \cite{nfl}:  
\begin{equation}  
\chi_{\lambda}^{\prime\prime}(\omega)=N_{F}  
{\omega\bar\tau_{\lambda}^{-1}\over{\bar\omega^{2}+  
\bar\tau_{\lambda}^{-2}}},  
\label{ten}  
\end{equation}  
where  
\begin{eqnarray}  
& &\bar\tau_{\lambda}^{-1}=  
\tau_{\lambda}^{*-1}+\alpha\sqrt{(\beta^{\prime} T)^{2}+\omega^{2}},  
\nonumber \\  
& &\bar\omega=\omega m^{*}(\omega)/m, \nonumber \\  
& &{\rm with}\ m^{*}(\omega)/m=1+{2\alpha\over{\pi}}  
{\rm ln}\left[{\omega_{c}\over{\sqrt{(\beta^{\prime}  
T)^{2}+\omega^{2}}}}\right].  
\label{eleven}  
\end{eqnarray}  
Here $N_{F}$ is the density of states for both spins at the Fermi level,
$m,m^{*}$ is the bare, renormalized electron mass, respectively,
$1/\tau_{\lambda}^{*}=1/\tau-1/\tau_{\lambda}$ is the  
channel-dependent impurity scattering rate reduced by vertex  
corrections\cite{ZC} and $\tau$ is the lifetime of the electrons.   
$\alpha, \beta^{\prime}$ and $\omega_{c}$ are constants determined by a  
fit to the electronic continuum.
The cut-off $\omega_{c}$ is essentially  
the band width, $\beta^{\prime}$ measures the interaction between the  
electrons and can vary only between 3.3 and 4.2. They both will be fixed  
as indicated in Table I. $\alpha \leq 1$ as long as $\beta^{\prime} \leq 3.7$ 
\cite{nfl}. 
In addition, $\alpha\cdot\beta^{\prime}$ determines the  
temperature dependence of the normal-state d.c. resistivity and should 
therefore be close to  
1.5. This means that there exist strong theoretical constraints for  
metallic samples. 
$1/\tau_{\lambda}^{*}$ is a free parameter to get an optimal  
representation of the continuum. This form for the Raman susceptibility  
provides an adequate description to the continuum in the normal  
state of both  
Y-123 and Bi-2212 (see Ref. \cite{nfl}). As the continuum depends only  
weakly on frequency it turns out that the choice of the parameters  
mildly influences the Fano line shape. The full Raman response  
measured  
in channel $\lambda$ for the interacting electron-phonon system reads  
\cite{DVZ1}:  
\begin{eqnarray}  
\chi_{\lambda,full}^{\prime\prime}(\omega)&=&  
{(\omega+\omega_{a})^{2}\over{(\omega^{2}- 
\hat\omega_{\lambda}^{2})^{2}+  
[2\omega_{\lambda}\Gamma_{\lambda}(\omega)]^{2}}}  
\Bigg\{\gamma_{\lambda}^{2}\chi_{\lambda}^{\prime\prime}(\omega)\left[  
(\omega- 
\omega_{a})^{2}+4\Gamma_{\lambda}^{i}\Gamma_{\lambda}(\omega)  
\left({\omega_{\lambda}\over{\omega+\omega_{a}}}\right)^{2}\right]  
\nonumber \\  
& &+4g_{p-p,\lambda}^{2}\Gamma_{\lambda}^{i}  
\left({\omega_{\lambda}\over{\omega+\omega_{a}}}\right)^{2}  
[1+\lambda(\omega)/\beta]^{2}\Bigg\}.  
\label{twelve}  
\end{eqnarray}  
Here, the first factor on the right hand side is the Lorentzian line shape of  
the phonon modified by the Fano interference and the second one  
describes the interplay between the scattering of light on phonons and  
electrons.   
The details depend on the relation between the coupling constants 
$g_{p-p,\lambda}$, which has been studied by Sherman and coworkers  
\cite{sherman}, $\gamma_{\lambda}$, and $g_{\lambda}$. The latter one  
describes the interaction of the phonon with the conduction electrons and  
can also be calculated directly from the band structure and the electric field
by averaging the  
coupling $g({\bf k})$ given in Eq. (\ref{eight}) over the Fermi surface.  
For $B_{1g}$ symmetry we obtain  
\begin{equation}  
g_{B_{1g}}^{2}=-{2\over{N}}\sum_{\bf k}\mid g({\bf k})\mid^{2}  
{\partial f(\epsilon({\bf k}))\over{\partial\epsilon({\bf k})}},  
\label{thirteen}  
\end{equation}  
where $f$ is the Fermi function and the factor of two accounts for both  
spin directions\cite{changes}. For convenience the dimensionless  
quantities $\lambda(\omega)$ and $\beta$ were introduced in Eq.  
(\ref{twelve}) which depend on the coupling constants as follows  
\begin{equation}  
\lambda(\omega_{B_{1g}})=  
{2g_{B_{1g}}^{2}\over{\omega_{B_{1g}}}}  
\label{fourteen}  
\end{equation}  
and  
\begin{equation}  
\beta={2g_{p-p,B_{1g}}g_{B_{1g}}\over{\gamma_{B_{1g}}
\omega_{B_{1g}}}\sqrt{N_{F}}}.  
\label{fifteen}  
\end{equation}  
$\lambda(\omega_{B_{1g}})$  
is the usual electron-phonon coupling constant. 
Here the explicit dependence of $\lambda(\omega)$ on $\omega$ through
$\chi_{B_{1g}}^{\prime}(\omega)$ \cite{DVZ1} has been dropped as  
the real part of the electronic response $\chi_{B_{1g}}^{\prime}(\omega)$ 
in the normal state turns out to be a smooth function of  
$\omega$ close to $\omega_{B_{1g}}$. 
Using these expressions  
we obtain the resonance frequency of the phonon  
$\hat\omega_{B_{1g}}$  
\begin{equation}  
\hat\omega_{B_{1g}}^{2}=\omega_{B_{1g}}^{2}[1- 
\lambda(\omega_{B_{1g}})]  
\label{sixteen}  
\end{equation}  
which depends only on the electron-phonon coupling. The position of  
the antiresonance $\omega_{a}^{2}$ can be given as  
\begin{equation}  
\omega_{a}^{2}=\omega_{B_{1g}}^{2}[1+\beta].  
\label{seventeen}  
\end{equation}  
Through $\beta$ the antiresonance frequency depends also on ratio 
of the photon-phonon coupling $g_{p-p,B_{1g}}$
to the projected photon-electron coupling $\gamma_{B_{1g}}$. The  
magnitude of these parameters is not important and enters as a  
multiplicative scale factor determining the absolute cross section in the  
same way as the density of states $N_{F}$. Finally, electron-phonon  
coupling leads to a larger line width  
\begin{equation}  
\Gamma_{B_{1g}}(\omega_{B_{1g}})=\Gamma^i_{B_{1g}}+  
g_{B_{1g}}^{2}\chi^{\prime\prime}_{B_{1g}}(\omega_{B_{1g}})/N_ 
{F},  
\label{eighteen}  
\end{equation}  
neglecting strong anisotropy of the electron-electron or electron-impurity
self energies.  
\section{Samples and Experimental}  
  
The Y-123 crystals were grown in   
BaZrO$_{3}$ \cite{grown} which has been shown to be completely inert   
and to facilitate the preparation of samples with a purity   
of better than 99.995\%. All crystals were postannealed   
in pure oxygen and quenched. Temperatures and   
oxygen partial pressures were adjusted according to the   
calibration of Lindemer et al.\cite{lind}. The resulting   
oxygen concentrations were approximately 6.0 
6.5, 6.93, and very close to   
7.0 for the samples we call antiferromagnetic (AF) insulating,  
under-doped, optimally   
doped and over-doped, respectively, in the following.   
The magnetically determined respective T$_{c}$ values and   
transition widths of the superconducting samples  
were 53.5 $K$ ($\Delta$T$_{c}$ = 3 $K$), 91.5   
($\Delta$T$_{c}$ =0.3 $K$), and 87.0 ($\Delta$T$_{c}$ = 1.0 $K$).   

The Bi-based samples, Bi$_{2}$Sr$_{2}$(Ca$_{1- 
x}$Y$_{x}$)Cu$_{2}$O$_{8}$   
(Bi-2212), were   
prepared in ZrO crucibles. In optimally doped crystals   
(without Y) the resistively measured T$_{c}$ was generally   
above 91 $K$ with $\Delta$T$_{c}$ $< 2 K$. If Ca$^{2+}$ is replaced  
by   
Y$^{3+}$ (see Fig. \ref{phonons})  
holes in the CuO$_{2}$ planes are filled in and $T_{c}$ is reduced.   
At the same time the environment of the planes   
becomes electrically asymmetric. The samples we used   
contained 38\% Y and 100\% Y, respectively. The superconducting  
sample (38\%)  
was well in the underdoped range   
of the phase diagram with a $T_{c}$ of 57 $K$ ($\Delta$T$_{c}$ = 5  
$K$).  
Bi-2212 is not stable at the stoichiometric composition and there is always  
excess Bi in the crystals (typically Bi$_{2.1}$ instead of Bi$_{2}$)  
\cite{GF96,revaz} which is found predominantly in the Ca (Y) position. In 
addition, a small amount of Sr may be replaced by Ca or Y. Therefore, the  
environment of the Cu-O plane is by far less ordered than Y-123. 
In general, it cannot be expected for Bi-2212 that a crystal  
quality comparable to the one of Y-123 is obtainable. By partially 
substituting Bi with Pb the modulation along the crystallographic 
b-axis can be changed or even suppressed completely \cite{Berger}. 
In the sample we used the Laue pattern still showed an indication 
of a twofold symmetry but it is reasonable to assume that the distortion 
is smaller than in Pb-free samples.
  
The experiments were performed in back-scattering   
geometry using a double monochromator with single-
channel detection and the resolution set at 8 cm$^{-1}$. This is sufficient  
as the total widths of the narrowest lines are roughly twice as large  
and add quadratically to the resolution. For   
excitation the Ar$^{+}$ line at 476 nm was selected. The   
maximal power was 4 mW in order to keep the laser-
induced heating below 15 $K$. The beam was focused to   
a spot of approximately $50 \times 150 \mu\rm{m}^{2}$. For the study of   
the excitations we were interested in the polarizations   
of the incoming and outgoing photons were always   
parallel to the planes. The coordinate system is locked   
to the Cu-O bonds with $x$ = [100], $x^{\prime}$ = [110], etc.   
All symmetries refer to a tetragonal pointgroup. $B_{1g}$   
phonon and continuum are projected out with $x^{\prime}y^{\prime}$   
polarization.   
  
\section{Results}  
  
\begin{figure}  
\vskip 0cm  
\epsfxsize=8cm  
\epsffile{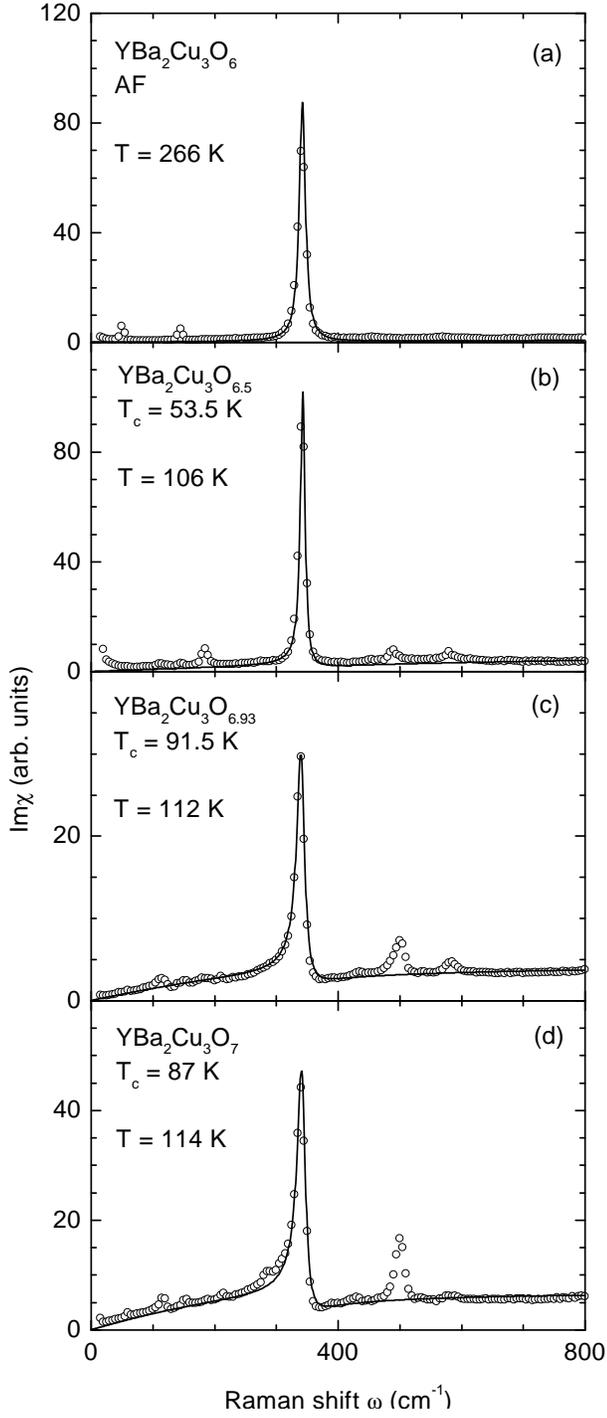}  
\vskip 1.0cm  
\caption{$B_{1g}$ spectra for differently doped  
YBa$_{2}$Cu$_{3}$O$_{6+x}$ as indicated. Shown are experimental  
data (after division by the Bose factor) and theoretical results from the  
crystal field model (Eqs. (\ref{twelve})--(\ref{eighteen})). The parameters  
used are defined in the text and compiled in Table I.}  
\label{Y123}  
\end{figure}

Results for Y-123 obtained at $B_{1g}$ symmetry  
$(x^{\prime}y^{\prime})$ are   
plotted in Fig. \ref{Y123}. All spectra are divided by the   
Bose-Einstein thermal function in order to get the   
response $Im\chi$ as given in Eq. (12). As a result of   
doping the shape of the $B_{1g}$ phonon at approximately   
330 cm$^{-1}$ changes considerably. For the AF and slightly doped  
sample   
it is narrow and close to a Lorentzian. When carriers are   
added the line broadens and becomes more asymmetric   
exhibiting a Fano-type dependence on frequency. The oscillator 
strength of the phonon decreases while  
the $B_{1g}$ continuum gains intensity\cite{katsu}. As a consequence the   
intensity ratio $I_{phonon}$ to $I_{continuum}$ decreases by a factor 
of at least 20. We emphasize that  
the low energy continuum vanishes almost completely in the AF crystal,  
hence supporting the interpretation in terms of charge excitations.  
  
\begin{figure*}  
\epsfxsize=8cm  
\epsffile{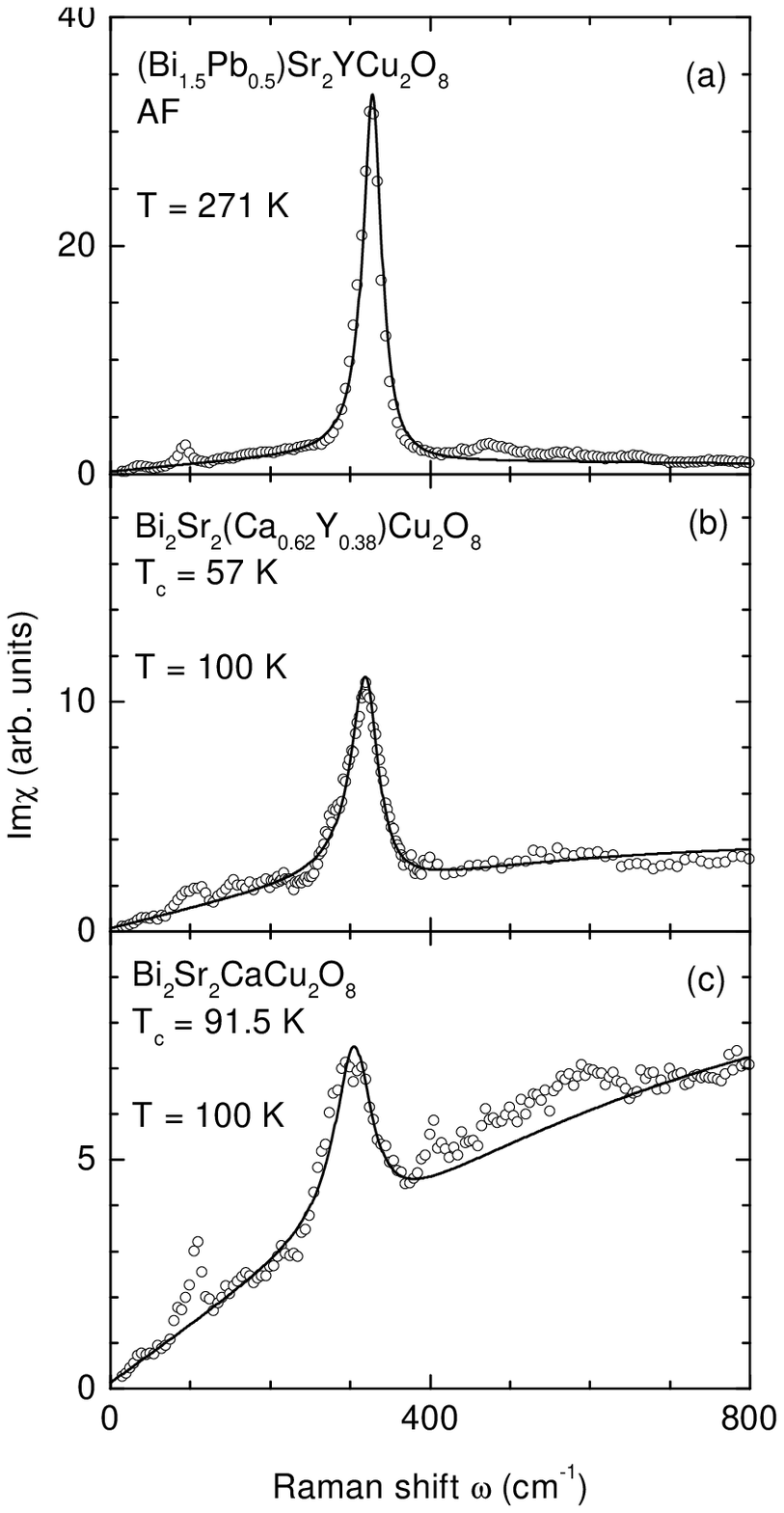}
\vskip 1.0cm  
\caption{$B_{1g}$ spectra for differently doped  
(Bi,Pb)Sr$_{2}$(Ca,Y)Cu$_{2}$O$_{8}$ as indicated. Shown are  
experimental data (after division by the Bose factor) and theoretical  
results from the crystal field model (Eqs. (\ref{twelve})--(\ref{eighteen})).  
The parameters used are defined in the text and compiled in Table I.}  
\label{Bi2212}  
\end{figure*} 
  
In optimally doped Bi-2212 (Fig. \ref{Bi2212} (c)) the $B_{1g}$   
phonon line is much weaker while the continuum has   
roughly the same cross section as comparable Y-123   
(Fig. \ref{Y123} (c)). The bigger line width comes at least in part   
from inhomogeneous broadening as explained above which manifests  
itself in the large intrinsic line width (see Table I).  
If Y is doped in for Ca a new line at approximately 315 cm$^{-1}$ starts to 
gain intensity very rapidly (Fig. \ref{Bi2212} (a) and (b)). A reminder of  
the line found in Y-free crystals is still seen as a shoulder on the   
left-hand side of the new one (Fig. \ref{Bi2212} (b)). This tells us that 
the Y is   
not completely homogeneously distributed in the underdoped sample. 
Most likely,  
the Y replaces big regions of the Ca planes, just like in Y-123 where  
upon oxygen doping clusters of longer and shorter fractions the 
chains build up depending on the conditions of 
preparation \cite{Janossy}.  So, in   
addition to changing the charge balance around the CuO$_{2}$ plane,  
defects are introduced in its vicinity  
which are seen in both the phononic   
and the electronic properties. Apparently the intrinsic
linewidth in the AF sample  
is somewhat smaller than in the metallic ones. This comes most likely  
from the additional doping with Pb which is used to suppress the   
superstructure of the BiO$_{2}$ layers and the related orthorhombic  
distortion of the unit cell. The integrated cross   
section is now as large as in AF Y-123 [compare Fig. \ref{Y123} (a) and  
Fig. \ref{Bi2212} (a)]. This is further independent evidence for the model 
proposed.  
  
\section{Discussion and Connection to Buckling}  
  
Excellent fits to the data on optimally-doped  
YBa$_{2}$Cu$_{3}$O$_{7-\delta}$ in both the superconducting and  
normal  
states were given in \cite{DVZ1}, which when interpreted supported  
evidence for $d_{x^{2}-y^{2}}$-pairing in this material. We have  
extended these results by examining normal state data of  
different samples and dopings. We note that a completely new set of Y- 
123 single crystals with superior quality (see section III) is used here. 

We find that once again very good fits to the data can be obtained on four  
and three differently doped samples of  
Y-123 and Bi-2212, respectively. For all fits given in Figs.   
\ref{Y123} and \ref{Bi2212}, the  
values used to calculate the susceptibilities of   
nested Fermi liquid theory and the phonon lineshapes are summarized in  
Table I. 

\begin{table}  
\begin{tabular}{|l|l|l|lllllll|l|lll|}  
Fig. & Sample & $T_{c}$ & Scale & $\alpha$ &  
${\tau_{B_{1g}}^{*}}^{-1}$ & $\Gamma_{B_{1g}}^{i}$ &  
$\omega_{B_{1g}}$ & $\omega_{a}$ & $\lambda$ &  
$\hat\omega_{B_{1g}}$ & $T$ & $\beta^{\prime}$ & $\omega_{c}$ \\   
 & & $[K]$ & & & [cm$^{-1}$] & [cm$^{-1}$] & [cm$^{-1}$] &  
[cm$^{-1}$] & &  
[cm$^{-1}$] & $[K]$ & & [cm$^{-1}$] \\ \hline  
3(a) & YBa$_{2}$Cu$_{3}$O$_{6}$ & AF &  4 & 0.55 & 3000 & 6.0 &  
342 & 345.5 &   
0.000585 & 342.1 & 266 & 3.3 & 12000 \\  
3(b) & YBa$_{2}$Cu$_{3}$O$_{6.5}$ & 53.5 & 22 & 0.55 & 3000 &  
4.0 & 347.5 &   
352.5 & 0.0257 & 343 & 106 & 3.3 & 12000 \\  
3(c) & YBa$_{2}$Cu$_{3}$O$_{6.93}$ & 91.5 & 20 & 0.95 & 1200 &  
6.5 & 348 &   
349.3 & 0.0426 & 340.5 & 112 & 3.3 & 12000 \\  
3(d) & YBa$_{2}$Cu$_{3}$O$_{7}$ & 87.0 & 30 & 0.75 & 900 & 6.0  
& 352 &   
352.5 & 0.056 & 342 & 114 & 3.3 & 12000 \\ \hline  
4(a) & (Bi$_{1.5}$Pb$_{0.5}$)Sr$_{2}$YCu$_{2}$O$_{8}$ & AF & 2.4  
& 0 & 400 &   
14 & 328.5 & 336.2 & 0.00213 & 328.15 & 271 & 3.3 & 12000 \\  
4(b) & Bi$_{2}$Sr$_{2}$Y$_{0.38}$Ca$_{0.62}$Cu$_{2}$O$_{8}$  
& 57.0 & 10 & 0.2 & 1200 &   
20 & 322 & 326 & 0.0124 & 320 & 100 & 3.3 & 12000 \\  
4(c) & Bi$_{2}$Sr$_{2}$CaCu$_{2}$O$_{8}$ & 91.5 & 20 & 0.1 &  
1600 &   
28 & 305.2 & 306.05 & 0.00131 & 305 & 100 & 3.3 & 12000 \\  
\end{tabular}  
\vskip 0.5cm  
\caption{Summary of fitting parameters used in Figs. \ref{Y123} and  
\ref{Bi2212}. $T_c$ is the superconducting critical temperature, 
$T$ the measuring temperature. AF indicates that the 
sample is antiferromagnetic and insulating.
The seven parameters in the third column are the fitted parameters. 
The actual Fano profile determines the parameters 
$\omega_{B_{1g}}$, $\omega_{a}$, and $\lambda$ 
which depend only weakly on $\alpha$, 
${\tau_{B_{1g}}^{*}}^{-1}$, $\beta^{\prime}$, 
and $\omega_{c}$ (see text). $\hat\omega_{B_{1g}}$ 
is calculated from $\omega_{B_{1g}}$ and $\lambda$. The last 
column are fixed parameters.  
A scale factor was multiplied to Eq. (\ref{twelve}) to account  
for the overall magnitude of the cross section. The scale factor
is proportional to the  
density of states at the Fermi energy $N_{F}$ and is, as expected, 
very small in the AF  
samples.}  
\end{table}  

Eight parameters ($\alpha, \tau_{B_{1g}}^{*}, \Gamma_{B_{1g}}^{i},
\omega_{B_{1g}}, \omega_{a}, \lambda, \hat\omega_{B_{1g}}$, and an overall
scale factor) are used to provide the fits in Figs. 
\ref{Y123} and \ref{Bi2212}. However, we do not have eight separate
degrees of freedom. $\alpha, \tau_{B_{1g}}^{*},$ and the overall scale
factor are chosen to reproduce the continuum for frequencies
away from the phonon. $\tau_{B_{1g}}^{*}$ determines the low frequency
part of the continuum while $\alpha$ determines the high frequency part.
The uncertainty in these parameters is less than 10\%. In addition, 
the parameters $\alpha$ and $\beta^{\prime}$ are in the narrow 
range imposed by the nFl model.\cite{nfl} 
Then $\hat\omega_{B_{1g}}$ is chosen to locate the center of the phonon
line, which can be accurately placed to within a fraction of a wavenumber 
(i.e. much better than the experimental resolution).
Next $\lambda$ is chosen to reproduce the local asymmetry around the phonon
line (Fano). This automatically fixes $\omega_{B_{1g}}$ via Eq.
(\ref{sixteen}) and sets the damping from the coupled electron-phonon system.
Then, the intrinsic line width $\Gamma_{B_{1g}}^{i}$ 
is set to account for any remaining
damping needed to account for the phonon linewidth. An  
indication of the stability is the almost constant intrinsic line width in Y-
123. Indeed these samples are comparable and the unit cell changes   
only slightly. On the other hand, 
it is not surprising that the intrinsic line width is not 
constant in Bi-2212 since the environment of the planes and even the  
total crystallographic structure changes considerably on doping. Finally,
$\omega_{a}$ is then chosen to set the overall intensity of the phonon
relative to the electronic continuum and to pinpoint the minimum of the
spectrum on the high frequency side of the phonon. In summary, each parameter
is independently chosen and the accuracy of the fit can be determined
by the agreement of the theory to the experimental background and the
phonon's position, intensity, asymmetry, linewidth, and local high frequency
minimum. Therefore we are confident that the parameters used are accurate
to within the 10\% level. Should further constraints be imposed (i.e. by
a more accurate theory for the continuum or the phonon-photon coupling
constant), it would be straightforward to adapt the theory accordingly.

We now can make a comparison to the crystal field model predictions for  
the electron-phonon coupling constant. We see from the fits that for the  
Y-123 compounds the electron-phonon coupling increases up to a  
factor of two with increasing doping. This can be understood in part by  
an increase in the density of states at the Fermi level with increased  
oxygen doping as the van Hove contribution moves closer to the Fermi  
level  (see Fig. \ref{band}).   
Our calculations using Eq. (\ref{fourteen}) yield the following  
values for $\lambda$ as a function of doping: for $\lambda=0.0321$ 
for $<n>=0.875$ (underdoped, Fig. 3b), 
$\lambda=0.0382$ for $<n>=0.85$ (optimally  
doped, Fig. 3c), 
and $\lambda=0.0482$ for $<n>=0.8$ (overdoped, Fig. 3d). Here we have   
taken $T=100K, \omega_{B_{1g}}=348$cm$^{-1}$,  
$M=16 m_{P}$, where $m_{P}$ is the proton mass, and used a  
value of the electric field which is $1.3$V/\AA\cite{Ladik}. 
We observe that these values of  
$\lambda$ are quite close to the values used to obtain the fits in Figs. 
\ref{Y123} (b)--(d). Given the values of $t^{\prime}/t, t$, and  
$\varepsilon$ are taken  
to be independent of doping, the agreement is quite good. The agreement  
could  
be refined once parameter choices for various levels of doping are  
determined via, e.g., fitting to experimentally observed Fermi surfaces.  
  
We can connect the value of the electric field used to account
for the observed buckling of the CuO$_{2}$ plane in Y-123. 
The electric field $E$ perpendicular to the CuO$_{2}$ plane   
results in an $A_{1g}$-type static distortion of the plane since   
the charges of oxygen and copper are different. It is   
very hard to perform a complete calculation as the   
electric field is different at the Cu and the O sites.   
Furthermore all the lattice forces have to be known. For   
an estimation of the buckling, however, a simplified   
model is sufficient\cite{Jorg} where Cu is pinned rigidly to the   
elementary cell and the oxygen moves in a harmonic   
potential being characterized by the frequency of the   
$A_{1g}$ phonon at $\omega_{A_{1g}}=435$ cm$^{-1}$\cite{frozen}.   
The restoring force at the buckling amplitude $\Delta z$ must balance the   
electric force acting on the oxygen with charge $q =   
- 1.75 e$. Thus, $qE = M\omega_{A_{1g}}^2 \Delta z$ 
must hold. With the experimental value $\Delta z = 0.24 \AA$\cite{Jorg},   
$E = 1.53 V/\AA$   
is obtained which is close both to the $1.3 V/\AA$ used for   
estimating the electron-phonon coupling strength and to  
the theoretically calculated number\cite{Ladik}. On the other   
hand, in the case of CuO$_{2}$ planes in a more symmetrical   
environment as in Bi-2212 and the infinite-layer   
compound CaCuO$_{2}$ the buckling, if it exists at all, is at   
least an order of magnitude smaller as found in   
structural studies\cite{Karp,karpnote}.   
  
At the same time, the electron-phonon coupling for optimally doped  
Bi-2212 (Fig. 4c) obtained from the fit is more  
than one order of magnitude smaller than for any of the metallic
Y-123 compounds  
and, as a result, the weight of the phonon is also reduced by at least  
the same amount.  
Indeed, in our model we would expect that the electron-phonon coupling  
is significantly smaller  
since the crystal electric field must be much weaker in this compound 
than in Y-123 (see Fig. \ref{phonons}). We attribute  
then the small electron-phonon coupling   
to a much smaller local asymmetry. A weaker spontaneous symmetry breaking is  
also possible. 
  
To test this idea we presented measurements of a sample of Bi-2212 which  
has been doped with small amounts of Y in place of Ca (Fig. 4b).  
Since here the valence of 
Y (+3) is different  
from that of Ca (+2), once again the mirror plane symmetry is broken and  
we would expect a much larger electron-phonon coupling than in the  
undoped compound.The fit to the $B_{1g}$ data  
using Eq. (\ref{twelve}) is given in Fig. \ref{Bi2212}(b) and yields 
$\lambda=0.0124$. The coupling constant is  
indeed increased over that of the optimally doped compound 
by an order of magnitude when Y is  
introduced  
and in accordance the intensity of the phonon line is also essentially  
enhanced. This trend continues in the AF sample Fig. 4(a), and the 
phonon cross section is now similar to that in Y-123. 

The intensity of the continuum 
in Bi-2212:Y does not approach zero as one would expect for an insulator. 
In-Y-123, too, the continuum is not vanish identically but is 
significantly smaller. At present, we do not know the origin of 
this intensity. It can be either photoexcited carriers as observed 
in semiconductors\cite{abstreiter}, an effect of remaining carriers, 
or coupling of
the phonon to the spin system\cite{Normand}. As a consequence 
the coupling $\lambda$ is not as small, though reduced considerably, 
as to be expected and actually observed in Y-123. This again 
gives strong support for the crystal field coupling model as the  
driving source of electron-phonon coupling.  

\section{Summary and Conclusions}  
  
In summary, studying Y-123 with different doping levels   
we have shown that the breaking of the local reflection   
symmetry through the CuO$_{2}$ planes due to the position   
and charges of the neighboring atoms results in an   
electric field across the planes which is sufficiently   
strong to produce both the observed buckling and the   
strong electron-phonon coupling being responsible for   
the Fano-type line shape of the respective $B_{1g}$ mode. In   
order to check this idea experiments were performed on   
Bi-2212 with and without Y doping. While the sample   
without Y shows very weak electron-phonon coupling,   
the interaction is enhanced by an order of magnitude   
and becomes comparable to the one in Y-123 if the local   
reflection symmetry is broken by replacing part of the   
Ca by Y. (For Bi-2212 and CaCuO$_{2}$ there is no   
experimental indication for spontaneous symmetry   
breaking.) As doping Bi-2212 by Y results in a change of  
T$_{c}$ from T$_{c}=91.5K$ to T$_{c}=57K$ accompanied by   
a large increase of the coupling $\lambda$, the  
$\omega=330$cm$^{-1}$ phonon can not play an important role  
for the superconductivity, in agreement with the conclusion  
of Savrasov and Andersen\cite{SA}.  
We conclude that the symmetry breaking due to  
the composition can explain both the buckling of the   
planes and the strong linear electron-phonon coupling   
in the two types of material. One should keep in mind   
that spontaneous symmetry breaking can occur also in   
other materials like La$_{2-x}$Sr$_{x}$CuO$_{4}$ and determine the   
properties. Further experimental and theoretical studies   
of CaCuO$_{2}$ would be useful.  
It is of interest in the present context that the so called infinite  
layer material CaCuO$_2$ does not show any sign of  
buckling\cite{Karp},  
consistently with the symmetric position of the CuO$_2$ plane.  
  
\section*{Acknowledgments}  
  
We gratefully acknowledge enlightening discussions with W. Pickett, 
O.K. Andersen, and B. Gy\"orffy. This work was supported by the 
Hungarian National Research Fund under  
Grant Nos. OTKA T020030, T016740, T02228/1996, T024005/1997.   
Acknowledgment (T.P.D.) is made  
to the Donors of The Petroleum Research Fund, administered by the   
American Chemical Society, for partial support of this research.  
A. Z. is grateful for the support by the Humboldt Foundation. The  
experiments have been supported partially by the Bayerische  
Forschungsstiftung through the ``Forschungsverbund Hochtemperatur- 
Supraleiter (FORSUPRA)''  
We are grateful to the BMBF for financial support via the program  
``Bilaterale Wissenschaftlich-Technische Zusammenarbeit'' under grant  
no. WTZ-UNG-052-96. One of us (T.P.D.) was partially supported by the  
American Hungarian Joint Fund No. 587.


\begin{references}
\bibitem{changes}  
We note that the definitions of $g$ and $\lambda$ given in Equations  
(13-14) are different   
than those used in Ref. \cite{DVZ1}.  
This is simply because in this paper we do not take a flat density  
of states.
\bibitem{karpnote}  
It remains to study why the electronic energy   
calculations provide a spontaneous symmetry breaking   
for CaCuO$_{2}$.
\end{references}
\end{document}